\documentclass[12pt, a4paper, twoside, reqno]{amsart}

\usepackage{geometry}

\usepackage[pdftex]{graphicx}

\usepackage[noadjust]{cite}

\usepackage{lipsum}

\usepackage{amssymb}
\usepackage{amsfonts}
\usepackage{amsmath}
\usepackage{latexsym}

\usepackage{caption}
\usepackage{subcaption}
\usepackage{float}
\usepackage{rotating}

\usepackage{enumerate}

\usepackage{tikz}

\usepackage[ruled, vlined, linesnumbered]{algorithm2e}
\usepackage{url}        %(1)    % These
\usepackage{hyperref}       %(2)    % packages
\usepackage{amsthm}     %(3)    % must be
\usepackage{multirow}
\usepackage{tabularx}
\makeatletter
\newcommand*\bigcdot{\mathpalette\bigcdot@{.75}}
\newcommand*\bigcdot@[2]{\mathbin{\vcenter{\hbox{\scalebox{#2}{$\m@th#1\bullet$}}}}}
\makeatother

\hypersetup{
    colorlinks = true,
    citecolor = black,
    filecolor = black,
    linkcolor = black,
    urlcolor = black
}
\def\BzeroCPG{contact $B_0$--VPG}

\def\diamond{K_4 - e}

\def\free{--free}

\def\drawingTreeT{
    \begin{tikzpicture}
    \vertexInCycleTypeOne{0}{0}{0}{0.75}{160}{v1};
    \vertexInCycleTypeOne{0}{0}{0}{0.75}{280}{v11};
    \vertexInCycleTypeOne{0.50}{1.25}{0}{0.75}{180}{v2};
    \vertexInCycleTypeOne{0.50}{1.25}{0}{0.75}{60}{v22};
    \vertexInCycleTypeZero{0.75}{0.75}{0}{120}{v3};
    \vertexInCycleTypeZero{1.5}{0.75}{0}{120}{v4};
    \vertexInCycleTypeOne{2.1}{1.15}{0}{0.75}{105}{v5};
    \vertexInCycleTypeOne{1.75}{0}{0}{0.75}{240}{v6};
    \vertexInCycleTypeOne{3}{1}{0}{0.75}{45}{v7};
    \vertexInCycleTypeOne{3}{1}{0}{0.75}{285}{v77};
    \vertexInCycleTypeZero{2.75}{-0.25}{0}{120}{v8};
    \vertexInCycleTypeOne{3.75}{0}{0}{0.75}{80}{v9};
    \vertexInCycleTypeOne{3.75}{0}{0}{0.75}{320}{v99};
    \vertexInCycleTypeOne{2.75}{-1.25}{0}{0.75}{330}{v0};
    \vertexInCycleTypeOne{2.75}{-1.25}{0}{0.75}{210}{v00};

    \edge{v1}{v3};
    \edge{v2}{v3};
    \edge{v3}{v4};
    \edge{v4}{v5};
    \edge{v4}{v6};
    \edge{v5}{v7};
    \edge{v6}{v8};
    \edge{v8}{v9};
    \edge{v8}{v0};
    \end{tikzpicture}
}

\def\drawingTypeOne{
    \begin{tikzpicture}
    \vertexInCycleTypeOne{0}{0}{0}{1}{90}{v};
    \vertexLabel[below]{v}{$v_i$};
    \end{tikzpicture}
}

\def\drawingTypeTwo{
    \begin{tikzpicture}
    \vertexInCycleTypeZero{0}{0}{1}{60}{v};
    \vertexLabel[below]{v}{$v_i$};
    \vertexInCycleTypeZero{0}{0}{1}{120}{u};
    \vertexLabel[below]{u}{$v_{i \pm 1}$};

    \edge{u}{v};
    \vertexInCycleTypeTwoKfour{0}{0}{1}{120}{60}{u}{v};
    \end{tikzpicture}
}

\def\drawingTypeThree{
    \begin{tikzpicture}
    \vertexInCycleTypeOne{0}{0}{1}{1}{60}{v};
    \vertexLabel[below]{v}{$v_i$};
    \vertexInCycleTypeZero{0}{0}{1}{120}{u};
    \vertexLabel[below]{u}{$v_{i \pm 1}$};

    \edge{u}{v};
    \vertexInCycleTypeTwoKfour{0}{0}{1}{120}{60}{u}{v};
    \end{tikzpicture}
}

\def\drawingTypeFour{
    \begin{tikzpicture}
    \vertexInCycleTypeFour{0}{0}{0}{1}{90}{v};
    \vertexLabel[right]{v}{$v_i$};
    \end{tikzpicture}
}

\def\drawingHzero{
    \begin{tikzpicture}
    \vertexInCycleTypeOne{0}{0}{0}{1}{90}{v};
    \vertexInCycleTypeOne{0}{0}{0}{1}{210}{v};
    \vertexInCycleTypeOne{0}{0}{0}{1}{330}{v};
    \end{tikzpicture}
}

\def\drawingKfive{
    \begin{tikzpicture}
    \complete{0}{0}{1}{5}{k5};
    \end{tikzpicture}
}

\def\drawingDiamond{
    \begin{tikzpicture}
    \vertex{0}{0.75}{a};
    \vertex{-0.4335}{0}{b};
    \vertex{0.4335}{0}{c};
    \vertex{0}{-0.75}{d};

    \edge{a}{b};
    \edge{a}{c};
    \edge{b}{c};
    \edge{b}{d};
    \edge{c}{d};
    \end{tikzpicture}
}

\def\drawingFone{
    \begin{tikzpicture}
    \vertexInCycleTypeFour{0}{0}{0}{1}{90}{v1};
    \vertexInCycleTypeOne{1}{0}{0}{1}{90}{v2};
    \vertexInCycleTypeOne{3}{0}{0}{1}{90}{v3};
    \vertexInCycleTypeFour{4}{0}{0}{1}{90}{v4};

    \edge{v1}{v2};
    \edge[dashed]{v2}{v3};
    \edge{v3}{v4};
    \end{tikzpicture}
}

\def\drawingFtwo{
    \begin{tikzpicture}
    \vertexInCycleTypeFour{0}{0}{1}{1}{90}{v1};
    \vertexInCycleTypeOne{0}{0}{1}{1}{60}{v2};
    \vertexInCycleTypeOne{0}{0}{1}{1}{30}{v3};
    \vertexInCycleTypeOne{0}{0}{1}{1}{120}{v4};

    \edge{v1}{v2};
    \edge{v2}{v3};
    \edge{v1}{v4};
    \draw[dashed] ([shift=(120:1)] 0,0) arc (120:390:1);
    \end{tikzpicture}
}

\def\drawingFthree{
    \begin{tikzpicture}
    \vertexInCycleTypeOne{0}{0}{1}{1}{90}{v1};
    \vertexInCycleTypeOne{0}{0}{1}{1}{60}{v2};
    \vertexInCycleTypeOne{0}{0}{1}{1}{30}{v3};
    \vertexInCycleTypeOne{0}{0}{1}{1}{120}{v4};
    \vertexInCycleTypeOne{0}{0}{1}{1}{150}{v5};

    \edge{v1}{v2};
    \edge{v2}{v3};
    \edge{v1}{v4};
    \edge{v4}{v5};
    \draw[dashed] ([shift=(150:1)] 0,0) arc (150:390:1);
    \end{tikzpicture}
}

\def\drawingFfourCaseA{
    \begin{tikzpicture}
    \vertexInCycleTypeFour{0}{0}{1}{0.75}{90}{v1};
    \vertexInCycleTypeOne{0}{0}{1}{0.75}{60}{v2};
    \vertexInCycleTypeOne{0}{0}{1}{0.75}{0}{v3};
    \vertexInCycleTypeZero{0}{0}{1}{330}{ve};
    \vertexInCycleTypeOne{0}{0}{1}{0.75}{300}{v4};
    \vertexInCycleTypeOne{0}{0}{1}{0.75}{120}{v5};

    \edge{v1}{v2};
    \draw[dashed] ([shift=(0:1)] 0,0) arc (0:60:1);
    \edge{v3}{ve};
    \edge{ve}{v4};
    \edge{v5}{v1};
    \draw[dashed] ([shift=(120:1)] 0,0) arc (120:300:1);
    \end{tikzpicture}
}

\def\drawingFfourCaseB{
    \begin{tikzpicture}
    \vertexInCycleTypeOne{0}{0}{1}{0.75}{120}{d1a};
    \vertexInCycleTypeFour{0}{0}{1}{0.75}{90}{d1b};
    \vertexInCycleTypeOne{0}{0}{1}{0.75}{60}{d1c};

    \vertexInCycleTypeOne{0}{0}{1}{0.75}{20}{s1a};
    \vertexInCycleTypeZero{0}{0}{1}{0}{s1b};
    \vertexInCycleTypeOne{0}{0}{1}{0.75}{340}{s1c};

    \vertexInCycleTypeOne{0}{0}{1}{0.75}{300}{d2a};
    \vertexInCycleTypeFour{0}{0}{1}{0.75}{270}{d2b};
    \vertexInCycleTypeOne{0}{0}{1}{0.75}{240}{d2c};

    \vertexInCycleTypeOne{0}{0}{1}{0.75}{200}{s2a};
    \vertexInCycleTypeZero{0}{0}{1}{180}{s2b};
    \vertexInCycleTypeOne{0}{0}{1}{0.75}{160}{s2c};

    \edge{d1b}{d1a};
    \edge{d1b}{d1c};
    \draw[dashed] ([shift=(20:1)] 0,0) arc (20:60:1);
    \edge{s1b}{s1a};
    \edge{s1b}{s1c};
    \draw[dashed] ([shift=(300:1)] 0,0) arc (300:340:1);
    \edge{d2b}{d2a};
    \edge{d2b}{d2c};
    \draw[dashed] ([shift=(200:1)] 0,0) arc (200:240:1);
    \edge{s2b}{s2a};
    \edge{s2b}{s2c};
    \draw[dashed] ([shift=(120:1)] 0,0) arc (120:160:1);
    \end{tikzpicture}
}

\def\drawingFfive{
    \begin{tikzpicture}
    \vertexInCycleTypeOne{0}{0}{1}{1}{60}{v2};
    \vertexInCycleTypeOne{0}{0}{1}{1}{30}{v3};
    \vertexInCycleTypeOne{0}{0}{1}{1}{120}{v4};
    \vertexInCycleTypeOne{0}{0}{1}{1}{150}{v5};

    \edge{v2}{v4};
    \vertexInCycleTypeTwoKfour{0}{0}{1}{120}{60}{v4}{v2};
    \edge{v2}{v3};
    \edge{v4}{v5};
    \draw[dashed] ([shift=(150:1)] 0,0) arc (150:390:1);
    \end{tikzpicture}
}

\def\drawingGridStair{
    \begin{tikzpicture}
    \gridLine{ 0.00}{-0.25}{ 0.00}{ 1.00}; % V
    \gridLine{ 0.00}{-0.25}{ 0.00}{-0.50}; % (K_4) v
    \gridLine{ 0.00}{-0.25}{-0.25}{-0.25}; % (K_4) h
    \gridLine{ 0.00}{-0.25}{ 0.25}{-0.25}; % (K_4) h

    \gridLine{-0.25}{ 1.00}{ 1.00}{ 1.00}; % H
    \gridLine{-0.25}{ 1.00}{-0.50}{ 1.00}; % (K_4) h
    \gridLine{-0.25}{ 1.00}{-0.25}{ 1.25}; % (K_4) v
    \gridLine{-0.25}{ 1.00}{-0.25}{ 0.75}; % (K_4) v

    \gridLine{ 1.00}{ 0.75}{ 1.00}{ 2.00}; % V
    \gridLine{ 1.00}{ 0.75}{ 1.00}{ 0.50}; % (K_4) v
    \gridLine{ 1.00}{ 0.75}{ 1.25}{ 0.75}; % (K_4) h
    \gridLine{ 1.00}{ 0.75}{ 0.75}{ 0.75}; % (K_4) h

    \gridLine{ 0.75}{ 2.00}{ 2.00}{ 2.00}; % H
    \gridLine{ 0.75}{ 2.00}{ 0.50}{ 2.00}; % (K_4) h
    \gridLine{ 0.75}{ 2.00}{ 0.75}{ 2.25}; % (K_4) v
    \gridLine{ 0.75}{ 2.00}{ 0.75}{ 1.75}; % (K_4) v

    \gridLine{ 2.00}{ 1.75}{ 2.00}{ 3.00}; % V
    \gridLine{ 2.00}{ 1.75}{ 2.00}{ 1.50}; % (K_4) v
    \gridLine{ 2.00}{ 1.75}{ 2.25}{ 1.75}; % (K_4) h
    \gridLine{ 2.00}{ 1.75}{ 1.75}{ 1.75}; % (K_4) h

    \gridLine{ 1.75}{ 3.00}{ 3.00}{ 3.00}; % H
    \gridLine{ 1.75}{ 3.00}{ 1.50}{ 3.00}; % (K_4) h
    \gridLine{ 1.75}{ 3.00}{ 1.75}{ 3.25}; % (K_4) v
    \gridLine{ 1.75}{ 3.00}{ 1.75}{ 2.75}; % (K_4) v

    \gridLine{ 3.00}{ 3.25}{ 3.00}{ 0.00}; % V
    \gridLine{ 3.00}{ 3.25}{ 3.00}{ 3.50}; % (K_4) v
    \gridLine{ 3.00}{ 3.25}{ 3.25}{ 3.25}; % (K_4) h
    \gridLine{ 3.00}{ 3.25}{ 2.75}{ 3.25}; % (K_4) h

    \gridLine{ 3.25}{ 0.00}{ 0.00}{ 0.00}; % H
    \gridLine{ 3.25}{ 0.00}{ 3.50}{ 0.00}; % (K_4) h
    \gridLine{ 3.25}{ 0.00}{ 3.25}{ 0.25}; % (K_4) v
    \gridLine{ 3.25}{ 0.00}{ 3.25}{-0.25}; % (K_4) v
    \end{tikzpicture}
}

\def\drawingGridStairCircle{
    \begin{tikzpicture}
    \draw ( 0.00, 0.25) -- ( 0.00, 0.75); % V

    \draw ( 0.25, 1.00) -- ( 0.75, 1.00); % H

    \draw ( 1.00, 1.25) -- ( 1.00, 1.75); % V

    \draw ( 1.25, 2.00) -- ( 1.75, 2.00); % H

    \draw ( 2.00, 2.25) -- ( 2.00, 2.75); % V

    \draw ( 2.25, 3.00) -- ( 2.75, 3.00); % H

    \draw ( 3.00, 2.75) -- ( 3.00, 0.25); % V

    \draw ( 2.75, 0.00) -- ( 1.75, 0.00); % H

    \draw ( 1.25, 0.00) -- ( 0.25, 0.00); % H

    \draw [dotted] ( 1.50,0.00) circle [radius=0.25];

    \draw [dashed] ( 0.00,0.00) circle [radius=0.25];

    \draw [dashed] ( 3.00,0.00) circle [radius=0.25];

    \draw [dashed] ( 0.00,0.00) circle [radius=0.25];

    \draw [dashed] ( 0.00,1.00) circle [radius=0.25];

    \draw [dashed] ( 1.00,1.00) circle [radius=0.25];

    \draw [dashed] ( 1.00,2.00) circle [radius=0.25];

    \draw [dashed] ( 2.00,2.00) circle [radius=0.25];

    \draw [dashed] ( 2.00,3.00) circle [radius=0.25];

    \draw [dashed] ( 3.00,3.00) circle [radius=0.25];

    \draw [dotted] (-3.50,1.50) circle [radius=0.75];
    \node at (-3.50,0.25) {$C$ even};
    \draw (-4.25,1.50) -- (-2.75,1.50); % H

    \draw [dotted] (-1.50,1.50) circle [radius=0.75];
    \node  at (-1.50,0.25) {$C$ odd};
    \draw [->, >=stealth] (-2.25,1.50) -- (-1.50,1.50); % HL
    \draw [->, >=stealth] (-0.75,1.50) -- (-1.50,1.50); % HR
    \gridLine{-1.50}{ 1.50}{-1.50}{ 2.00}; % (K_4) v
    \gridLine{-1.50}{ 1.50}{-1.50}{ 1.00}; % (K_4) v

    \draw [dashed] ( 4.50,1.50) circle [radius=0.75];
    \node  at ( 4.50,0.25) {$(a)$};
    \draw [->, >=stealth] ( 5.25,1.50) -- ( 4.25,1.50); % H
    \draw [->, >=stealth] ( 4.50,0.75) -- ( 4.50,1.50); % V
    \gridLine{ 4.25}{ 1.50}{ 4.25}{ 1.25}; % (K_4) v
    \gridLine{ 4.25}{ 1.50}{ 4.25}{ 1.75}; % (K_4) v
    \gridLine{ 4.25}{ 1.50}{ 4.00}{ 1.50}; % (K_4) h

    \draw [dashed] ( 6.50,1.50) circle [radius=0.75];
    \node  at ( 6.50,0.25) {$(b)$};
    \draw [->, >=stealth] ( 7.25,1.50) -- ( 6.50,1.50); % H
    \draw [->, >=stealth] ( 6.50,0.75) -- ( 6.50,1.75); % V
    \gridLine{ 6.50}{ 1.75}{ 6.50}{ 2.00}; % (K_4) v
    \gridLine{ 6.50}{ 1.75}{ 6.25}{ 1.75}; % (K_4) h
    \gridLine{ 6.50}{ 1.75}{ 6.75}{ 1.75}; % (K_4) h

    \draw [dashed] ( 8.50,1.50) circle [radius=0.75];
    \node  at ( 8.50,0.25) {$(c)$};
    \draw [->, >=stealth] ( 9.25,1.50) -- ( 8.50,1.50); % H
    \draw [->, >=stealth] ( 8.50,0.75) -- ( 8.50,1.50); % V
    \gridLine{ 8.50}{ 1.50}{ 8.50}{ 2.00}; % (K_4) v
    \gridLine{ 8.50}{ 1.50}{ 8.00}{ 1.50}; % (K_4) h

    \end{tikzpicture}
}

\def\drawingOrientation{
    \begin{tikzpicture}

    \vertexInCycleTypeFour[lightgray]{0}{0}{1}{1}{90}{v4};
    \vertexInCycleTypeZero{0}{0}{1}{90}{v1};
    \vertexInCycleTypeOne{0}{0}{1}{1}{60}{v2};
    \vertexInCycleTypeOne{0}{0}{1}{1}{0}{v3};
    \vertexInCycleTypeZero{0}{0}{1}{330}{ve};

    \edge{v1}{v2};
    \edgeCurved[dashed,->,>=stealth,ultra thick]{v3}{v2}{90}{330};
    \edge[->,>=stealth,ultra thick]{ve}{v3};

    \vertexInCycleTypeFour{5}{0}{1}{1}{90}{w1};
    \vertexInCycleTypeOne[lightgray]{5}{0}{1}{1}{60}{w4};
    \vertexInCycleTypeZero{5}{0}{1}{60}{w2};
    \vertexInCycleTypeOne[lightgray]{5}{0}{1}{1}{120}{w5};
    \vertexInCycleTypeZero{5}{0}{1}{120}{w3};

    \edge[<-,>=stealth,ultra thick]{w1}{w2};
    \edge[->,>=stealth,ultra thick]{w3}{w1};

    \vertexInCycleTypeOne{8.5}{0}{1}{1}{60}{v};
    \vertexInCycleTypeZero{8.5}{0}{1}{120}{u};
    \vertexInCycleTypeOne[lightgray]{8.5}{0}{1}{1}{30}{x};
    \vertexInCycleTypeZero{8.5}{0}{1}{30}{z};

    \edge{u}{v};
    \vertexInCycleTypeTwoKfour{8.5}{0}{1}{120}{60}{u}{v};
    \edge[->,>=stealth,ultra thick]{z}{v};

    \end{tikzpicture}
}

\newcommand{\gridLine}[4]{
    \draw[<->, >=stealth] (#1, #2) -- (#3, #4);
}

\newcommand{\vertex}[4][black]{
    \draw[#1, fill=#1, inner sep=0pt] (#2, #3) circle (0.075) node(#4){};
}

\newcommand{\vertexLabel}[3][above]{
    \path (#2) node[#1]{#3};
}
\newcommand{\edge}[3][]{
	\draw[#1] (#2) -- (#3);
}

\newcommand{\edgeCurved}[5][]{
	\draw[#1] (#2) to[out=#4, in=#5] (#3);
}

\newcommand{\vertexRegularPolygon}[7][]{
	\vertex[#1]
	{ { (#2) + (#4) * -cos(deg(6.283*(#6 - 1)/(#5)-(1.571+3.142/(#5)))) } }
	{ { (#3) + (#4) * sin(deg(6.283*(#6 - 1)/(#5)-(1.571+3.142/(#5)))) } }
	{#7};
}
\newcommand{\complete}[6][]{
	\foreach \completeI in {1,...,#5} {
		\vertexRegularPolygon[#1]
						{#2}{#3}
						{#4}
						{#5}
						{\completeI}
						{#6\completeI}
						{};
	}

	\foreach \completeI in {2,...,#5} {
		\pgfmathtruncatemacro{\completeIminusOne}{\completeI - 1};

		\foreach \completeJ in {1,...,\completeIminusOne} {
			\edge[#1]{#6\completeI}{#6\completeJ};
		}
	}
}
\newcommand{\fatSpider}[6][]{
	\foreach \fatSpiderI in {1,...,#5} {
		\vertexRegularPolygon[#1]
						{#2}{#3}
						{#4}
						{#5}
						{\fatSpiderI}
						{#6c\fatSpiderI}
						{};

		\vertexRegularPolygon[#1]
						{#2}{#3}
						{2*(#4)}
						{#5}
						{\fatSpiderI}
						{#6s\fatSpiderI}
						{};
	}

	\foreach \fatSpiderI in {2,...,#5} {
		\pgfmathtruncatemacro{\fatSpiderIminusOne}{\fatSpiderI - 1};

		\foreach \fatSpiderJ in {1,...,\fatSpiderIminusOne} {
			\edge[#1]{#6c\fatSpiderI}{#6c\fatSpiderJ};
		}
	}

	\foreach \fatSpiderI in {1,...,#5} {
		\foreach \fatSpiderJ in {1,...,#5} {
			\ifthenelse
			{\equal{\fatSpiderI}{\fatSpiderJ}}
			{}
			{\edge[#1]{#6s\fatSpiderI}{#6c\fatSpiderJ};};
		}
	}
}
\newcommand{\vertexInCycleTypeZero}[6][black]{
    \vertex[#1]
    { { (#2) + (#4) * cos(#5) } } % x
    { { (#3) + (#4) * sin(#5) } } % y
    {#6}; % name
}

\newcommand{\vertexInCycleTypeOne}[7][black]{
    \vertexInCycleTypeZero[#1]{#2}{#3}{#4}{#6}{#7};

    \vertex[#1]
    { { (#2)+(#4)*cos(#6) + (sqrt(3.0)/3.0*(#5)) * cos((#6)+30.0) } } % x
    { { (#3)+(#4)*sin(#6) + (sqrt(3.0)/3.0*(#5)) * sin((#6)+30.0) } } % y
    {#7a}; % name
    \vertex[#1]
    { { (#2) + ((#4)+(#5)) * cos(#6) } } % x
    { { (#3) + ((#4)+(#5)) * sin(#6) } } % y
    {#7b}; % name
    \vertex[#1]
    { { (#2)+(#4)*cos(#6) + (sqrt(3.0)/3.0*(#5)) * cos((#6)-30.0) } } % x
    { { (#3)+(#4)*sin(#6) + (sqrt(3.0)/3.0*(#5)) * sin((#6)-30.0) } } % y
    {#7c}; % name

    \edge[color=#1]{#7}{#7a};
    \edge[color=#1]{#7}{#7b};
    \edge[color=#1]{#7}{#7c};
    \edge[color=#1]{#7a}{#7b};
    \edge[color=#1]{#7a}{#7c};
    \edge[color=#1]{#7b}{#7c};
}

\newcommand{\vertexInCycleTypeTwoKfour}[8][black]{
    \vertex[#1]
     { { (#2)+(#4)*cos(#5) + 0.5*sqrt((#4)*(cos(#5)-cos(#6))*(#4)*(cos(#5)-cos(#6)) + (#4)*(sin(#5)-sin(#6))*(#4)*(sin(#5)-sin(#6))) * cos(asin(((#4)*(cos(#6)-cos(#5)))/(sqrt((#4)*(cos(#5)-cos(#6))*(#4)*(cos(#5)-cos(#6)) + (#4)*(sin(#5)-sin(#6))*(#4)*(sin(#5)-sin(#6))))) - 30.0) } } % x
     { { (#3)+(#4)*sin(#5) + 0.5*sqrt((#4)*(cos(#5)-cos(#6))*(#4)*(cos(#5)-cos(#6)) + (#4)*(sin(#5)-sin(#6))*(#4)*(sin(#5)-sin(#6))) * sin(asin(((#4)*(cos(#6)-cos(#5)))/(sqrt((#4)*(cos(#5)-cos(#6))*(#4)*(cos(#5)-cos(#6)) + (#4)*(sin(#5)-sin(#6))*(#4)*(sin(#5)-sin(#6))))) - 30.0) } } % y
     {#7#8a}; % name

    \vertex[#1]
     { { (#2)+(#4)*cos(#6) + 0.5*sqrt((#4)*(cos(#5)-cos(#6))*(#4)*(cos(#5)-cos(#6)) + (#4)*(sin(#5)-sin(#6))*(#4)*(sin(#5)-sin(#6))) * cos(asin(((#4)*(cos(#6)-cos(#5)))/(sqrt((#4)*(cos(#5)-cos(#6))*(#4)*(cos(#5)-cos(#6)) + (#4)*(sin(#5)-sin(#6))*(#4)*(sin(#5)-sin(#6))))) + 30.0) } } % x
     { { (#3)+(#4)*sin(#6) + 0.5*sqrt((#4)*(cos(#5)-cos(#6))*(#4)*(cos(#5)-cos(#6)) + (#4)*(sin(#5)-sin(#6))*(#4)*(sin(#5)-sin(#6))) * sin(asin(((#4)*(cos(#6)-cos(#5)))/(sqrt((#4)*(cos(#5)-cos(#6))*(#4)*(cos(#5)-cos(#6)) + (#4)*(sin(#5)-sin(#6))*(#4)*(sin(#5)-sin(#6))))) + 30.0) } } % y
     {#7#8b}; % name

     \edge[color=#1]{#7}{#7#8a};
     \edge[color=#1]{#7}{#7#8b};
     \edge[color=#1]{#8}{#7#8a};
     \edge[color=#1]{#8}{#7#8b};
     \edge[color=#1]{#7#8a}{#7#8b};
}

\newcommand{\vertexInCycleTypeFour}[7][black]{
    \vertexInCycleTypeOne[#1]{#2}{#3}{#4}{#5}{#6}{#7};

    \vertex[#1]
    { { (#2)+(#4)*cos(#6) - (sqrt(3.0)/3.0*(#5)) * cos((#6)+30.0) } } % x
    { { (#3)+(#4)*sin(#6) - (sqrt(3.0)/3.0*(#5)) * sin((#6)+30.0) } } % y
    {#7x}; % name
    \vertex[#1]
    { { (#2) + ((#4)-(#5)) * cos(#6) } } % x
    { { (#3) + ((#4)-(#5)) * sin(#6) } } % y
    {#7y}; % name
    \vertex[#1]
    { { (#2)+(#4)*cos(#6) - (sqrt(3.0)/3.0*(#5)) * cos((#6)-30.0) } } % x
    { { (#3)+(#4)*sin(#6) - (sqrt(3.0)/3.0*(#5)) * sin((#6)-30.0) } } % y
    {#7z}; % name

    \edge[color=#1]{#7}{#7x};
    \edge[color=#1]{#7}{#7y};
    \edge[color=#1]{#7}{#7z};
    \edge[color=#1]{#7x}{#7y};
    \edge[color=#1]{#7x}{#7z};
    \edge[color=#1]{#7y}{#7z};
}

\newtheoremstyle{TheoremNum}
        {\topsep}{\topsep}                  % space between body and thm
        {\itshape}                          % thm body font
        {}                              % indent amount (empty = no indent)
        {\bfseries}                         % thm head font
        {.}                             % punctuation after thm head
        { }                             % space after thm head
        {\thmname{#1}\thmnote{ \bfseries #3}}   % thm head spec

\theoremstyle{TheoremNum}

\theoremstyle{plain}
\newtheorem{lemma}{Lemma}[section]

\newtheorem{theorem}[lemma]{Theorem}

\newtheorem{corollary}[lemma]{Corollary}

\theoremstyle{remark}

\newtheorem{remark}[lemma]{Remark}

\theoremstyle{definition}

\begin{document}

%---------------------------------------------------------------------------------------------------
%---------------------------------------------------------------------------------------------------
% Title and abstract
\def\UBA{Universidad de Buenos Aires}
\def\FCEN{Facultad de Ciencias Exactas y Naturales}
\def\DC{Departamento de Computaci\'on}
\def\BuenosAiresArgentina{Buenos Aires, Argentina}
\def\ICC{CONICET-\UBA. Instituto de Investigaci\'on en Ciencias de la Computaci\'on (ICC)}

\title{Characterising circular-arc contact $B_0$--VPG graphs}

\author[F. Bonomo]{Flavia Bonomo-Braberman}
\address{\UBA. \FCEN. \DC. \BuenosAiresArgentina. / \ICC. \BuenosAiresArgentina.}
\email{fbonomo@dc.uba.ar}

\author[E. Galby]{Esther Galby}
\address{University of Fribourg. Department of Informatics. Decision Support \& Operations Research. Fribourg, Switzerland}
\email{esther.galby@unifr.ch}

\author[C. L. Gonzalez]{Carolina Luc{\'i}a Gonzalez}
\address{\ICC. \BuenosAiresArgentina.}
\email{cgonzalez@dc.uba.ar}

\begin{abstract}
A \emph{\BzeroCPG\ graph} is a graph for which there exists a collection of nontrivial pairwise interiorly disjoint horizontal and vertical segments in one-to-one correspondence with its vertex set such that two vertices are adjacent if and only if the corresponding segments touch. It was shown in~\cite{D-G-M-R-cpg} that {\sc Recognition} is $\mathsf{NP}$-complete for \BzeroCPG\ graphs. In this paper we present a minimal forbidden induced subgraph characterisation of \BzeroCPG\ graphs within the class of circular-arc graphs and provide a polynomial-time algorithm for recognising these graphs.
\end{abstract}


\maketitle

%---------------------------------------------------------------------------------------------------
%---------------------------------------------------------------------------------------------------
% Introduction and preliminaries
\section{Introduction}
\label{sec:introduction}

Intersection graphs of various types of objects have been extensively studied in the last sixty years (see for example~\cite{M-M-intersection}). In~\cite{asinowski}, Asinowski et al. introduced the class of \textit{Vertex intersection graphs of Paths on a Grid} (\textit{VPG graphs} for short) which consists of those graphs whose vertices may be representated by paths on a grid in such a way that two vertices are adjacent if and only if the corresponding paths intersect on at least one grid-point.  
It is not difficult to see that the class of VPG graphs coincides with that of string graphs~\cite{E-E-T-intersec}, that is, intersection graphs of curves in the plane (see~\cite{asinowski}).

A natural restriction which was forthwith considered consists in limiting the number of \textit{bends} (i.e. $90$ degrees turns at
a grid-point) that the paths may have: a graph is a \textit{$B_k$-VPG graph}, for some integer $k \geq 0$, if one can assign a path on a grid having at most $k$ \textit{bends} to each vertex such that two vertices are adjacent if and only if the corresponding paths intersect on at least one grid-point. Since
their introduction, $B_k$-VPG graphs have received much attention (see for instance \cite{ABM-VPG-gc, asinowski, chaplick, chaplick12, cohen1,cohen2, Felsner, francis, Golumbic-Ries, gonca,Knau-epg-planar}).

A notion closely related to intersection graphs is that of \textit{contact graphs}. Such graphs can be seen as a special type
of intersection graphs of geometrical objects in which these objects are pairwise interiorly disjoint. Similarly to intersections graphs, contact graphs of various types of objects have been extensively studied in the literature (see for instance \cite{A-F-cpg,A-F-cpg-jour, castro, Fraysseix1,Fraysseix2,Felsner,Hlineny,
Hlineny1bis, Hlineny1}). In this paper, we are interested in the contact counterpart of VPG graphs, namely \textit{Contact graphs of Paths on a Grid} (\textit{contact VPG graphs} for short, also known as \emph{CPG graphs}) which are defined as follows. A graph $G$ is a \textit{contact VPG graph} if the vertices of $G$ can be represented by a family of nontrivial and pairwise interiorly disjoint paths on a grid in such a way that two vertices are adjacent in $G$ if and only if the corresponding paths touch, that is, share a grid-point which is an endpoint of at least one of the two paths. Note that this class is hereditary, i.e., closed under vertex deletion. Similarly to VPG graphs, a \emph{contact $B_k$-VPG graph} is a contact VPG graph admitting a representation in which each path has at most $k$ bends. Clearly, any contact $B_k$-VPG graph is also a $B_k$-VPG graph.

In this paper, we focus solely on \BzeroCPG\ graphs. It was shown in \cite{D-G-M-R-cpg,CPGhardness} that recognising the class of \BzeroCPG\ graphs is $\mathsf{NP}$-complete, and the complete list of minimal forbidden induced subgraphs for the class is not yet known. Nevertheless, characterisations of \BzeroCPG\ graphs by minimal forbidden induced subgraphs are known when restricted to some graph classes such as chordal, $P_5$-free, $P_4$-tidy, tree-cographs \cite{BMRR-ISCO18,chordalContactB0VPG}; furthermore, most of those characterisations lead to polynomial-time recognition algorithms within the class. It is also known that every bipartite planar graph is \BzeroCPG~\cite{Fraysseix2}. We here provide a characterisation of \BzeroCPG\ graphs by minimal forbidden induced subgraphs within the class of circular-arc graphs, i.e., intersection graphs of arcs of a circle~\cite{H-D-K-circ-arc,Klee-circ-arc} (see Section \ref{sec:characterization}), and a polynomial-time recognition algorithm for this class (see Section \ref{sec:algorithm}). We first give some terminology in Section \ref{sec:definitions} and some preliminary results in Section \ref{sec:preliminaries}. 

\section{Basic definitions}
\label{sec:definitions}

Let $G$ be a finite, simple and undirected graph with vertex set $V(G)$ and edge set $E(G)$.
For any $W\subseteq V(G)$, we denote by $G[W]$ the subgraph of $G$ induced by $W$.

Let $N(v)$ be the set of neighbours of $v \in V(G)$ and $N[v] = N(v) \cup \{v\}$. A vertex is \emph{simplicial} if its neighbours are pairwise adjacent. If $H$ is an induced subgraph of $G$ and $v$ a vertex of $G$, we denote by $N_H(v)$ the set $N(v) \cap V(H)$ and by $G-H$ the graph $G[V(G)-V(H)]$.

Let $v$ and $w$ be two vertices of $G$. The graph $G'$ obtained by the \emph{contraction} of $v$ and $w$ has vertex set $V(G) - \{w\}$ and edge set ($E(G) - \{wz : z\in N(w)\}) \cup \{vz : z\in N(w), z \neq v\}$.

Let $A,B\subseteq V(G)$. We say that $A$ \emph{is complete to} $B$ if every vertex of $A$ is adjacent to every vertex of $B$; and $A$ \emph{is anticomplete to} $B$ if no vertex of $A$ is adjacent to a vertex of $B$. A \emph{stable set} is a set of pairwise nonadjacent vertices. A graph $G$ is \emph{bipartite} if $V(G)$ can be partitioned into two stable sets $V_1$, $V_2$; and $G$ is \emph{complete bipartite} if $V_1$ is complete to $V_2$. We denote by $K_{r,s}$ the complete bipartite graph with $|V_1|=r$ and $|V_2|=s$. The \emph{claw} is the complete bipartite graph $K_{1,3}$. The \emph{bipartite claw} is the graph arising by subdividing the three edges of the claw.

We denote by $K_r$ ($r \geq 0$) the complete graph on $r$ vertices; $K_3$ will be also called a \emph{triangle}. A \emph{clique} in $G$ is a subset of vertices which induces a complete subgraph. A \emph{diamond}, also known as $\diamond$, is the graph obtained from $K_4$ by removing exactly one edge.

Let $P$ be a path in $G$. We denote by $P= v_1\dots v_k$ the fact that $V(P)=\{v_1,\dots,v_k\}$ and $v_i$ is adjacent to $v_{i+1}$ for $1 \leq i \leq k-1$. Vertices $v_1$ and $v_k$ are the \emph{extreme vertices} of $P$, while vertices in $V(P)-\{v_1,v_k\}$ are the \emph{internal vertices} of $P$. Similarly, let $C$ be a cycle in $G$. We denote by $C= v_1\dots v_k$ the fact that $V(C)=\{v_1,\dots,v_k\}$ and $v_i$ is adjacent to $v_{i+1}$ for $1 \leq i \leq k$, where indexes should be understood modulo $k$ (throughout the paper). An edge joining two nonconsecutive vertices of a path or a cycle in a graph is called a \emph{chord}. An \emph{induced path} is a chordless path in a graph. Likewise, an \emph{induced cycle} is a chordless cycle in a graph. A \emph{hole} is an induced cycle of length at least $4$. A graph is \emph{chordal} if it does not contain any hole. A hole is \emph{odd} if it has an odd number of vertices, and \emph{even}, otherwise.

Let $G$ and $H$ be two graphs. We say that $G$ is \emph{$H$\free\ }if $G$ does not contain an induced subgraph isomorphic to $H$. If $\mathcal H$ is a family of graphs, we say that $G$ is \emph{$\mathcal H$\free\ }if $G$ is $H$\free\ for every $H\in\mathcal H$.

A graph $G$ is a \emph{circular-arc} graph if it is the intersection graph of a set $\mathcal{S}$ of arcs on a circle, i.e., if there exists a one-to-one correspondence between the vertices of $G$ and the arcs of $\mathcal{S}$ such that two vertices of $G$ are adjacent if and only if the corresponding arcs in $\mathcal{S}$ intersect. Circular-arc graphs can be recognised in linear time~\cite{MC-circ-arc}, and have been characterised recently by a family of \emph{obstacles}~\cite{Hell-circ-arc}. Previously, partial characterisations by minimal forbidden induced subgraphs were presented in~\cite{TrotterMoore76} and~\cite{B-D-G-S-circ-arc}.

\section{Preliminary results}
\label{sec:preliminaries}

We first introduce some known families of minimal forbidden induced subgraphs for the class of \BzeroCPG\ graphs.

%------------------------------------------------
\begin{figure}[h]
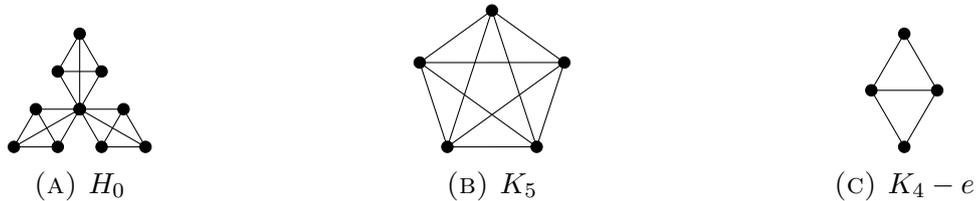

    \centering
    \begin{subfigure}[t]{0.3\textwidth}
        \centering
        \drawingHzero
        \caption{$H_0$}
    \label{fig:H0}
    \end{subfigure}%
    \hfill
    \begin{subfigure}[t]{0.3\textwidth}
        \centering
        \drawingKfive
        \caption{$K_5$}
    \label{fig:K5}
    \end{subfigure}%
    \hfill
    \begin{subfigure}[t]{0.3\textwidth}
        \centering
        \drawingDiamond
        \caption{$\diamond$}
    \label{fig:diamond}
    \end{subfigure}
    \caption{Some forbidden induced subgraphs for \BzeroCPG\ graphs.}
    \label{fig:H0K5D}
\end{figure}

\begin{lemma}\cite{D-G-M-R-cpg,BMRR-ISCO18}\label{lemma:H0K5D}
$H_0$, $K_5$ and $\diamond$ are not \BzeroCPG.
\end{lemma}

\begin{figure}[h]
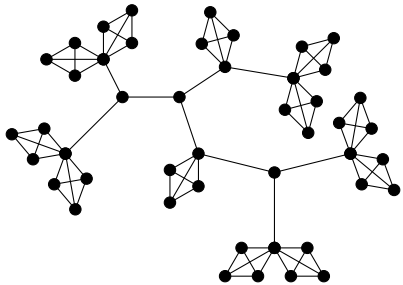

    \centering
    \drawingTreeT
    \caption{An example of a graph in $\mathcal{T}$.}
    \label{fig:tree}
\end{figure}

%-------------------------------------------------------
Let $\mathcal{T}$~\cite{BMRR-ISCO18} be the family of graphs containing $H_0$ (see Figure~\ref{fig:H0K5D}) as well as all graphs that can be partitioned into a nontrivial tree $T$ of maximum degree at most three and the disjoint union of triangles, in such a way that each triangle is complete to a vertex $v$ of $T$ and anticomplete to $T - \{v\}$, every leaf $v$ of $T$ is complete to exactly two triangles, every vertex $v$ of degree two in $T$ is complete to exactly one triangle, and vertices of degree three in $T$ have no neighbours outside $T$ (see Figure~\ref{fig:tree}).

\begin{theorem}\cite{BMRR-ISCO18}\label{thm:contact_chordal}
Let $G$ be a chordal graph. Then, $G$ is a \BzeroCPG\ graph if and only if $G$ is $\{\mathcal{T},K_5,\diamond\}$\free.
\end{theorem}

%------------------------------------------------
Let $\mathcal{F}_1$ be the family of graphs in $\mathcal{T}$ such that the tree $T$ is a path.

\begin{figure}[h]
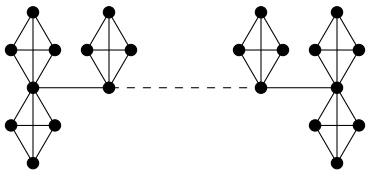

    \centering
    \drawingFone
    \caption{$\mathcal{F}_1$: The family of graphs in $\mathcal{T}$ such that the tree $T$ is a path.}
    \label{fig:F1}
\end{figure}

\begin{lemma}\label{lemma:F1}
The graphs in $\mathcal{F}_1$ are not \BzeroCPG.
\begin{proof}
$\mathcal{F}_1$ is a subfamily of $\mathcal{T}$ and it was shown in \cite{BMRR-ISCO18} that no graph in $\mathcal{T}$ is \BzeroCPG.
\end{proof}
\end{lemma}

It is easy to see that $\{H_0\} \cup \mathcal{F}_1$ is the family of graphs of $\mathcal{T}$ that do not contain a bipartite claw as induced subgraph (if a graph in $\mathcal{T}$ contains an induced bipartite claw then the tree $T$ must contain a vertex of degree three, and conversely, if $T$ contains a vertex of degree three then the graph contains an induced bipartite claw). Since the bipartite claw is not a circular-arc graph~\cite{TrotterMoore76}, we have the following corollary.

\begin{corollary}\label{cor:contact_chordal_bc-free}
Let $G$ be a (bipartite claw)\free\ chordal graph. Then, $G$ is a \BzeroCPG\ graph if and only if $G$ is $\{\mathcal{F}_1,H_0,K_5,\diamond\}$\free.
\end{corollary}

The next result easily follows.

\begin{corollary}\label{cor:contact_chordalCA}
Let $G$ be a chordal circular-arc graph. Then, $G$ is a \BzeroCPG\ graph if and only if $G$ is $\{\mathcal{F}_1,H_0,K_5,\diamond\}$\free.
\end{corollary}

%------------------------------------------------
In~\cite{B-D-G-S-circ-arc}, circular-arc graphs are characterised within some graph classes including, among others, the class of diamond\free\ graphs. The following is a straightforward corollary of Theorem~16 in~\cite{B-D-G-S-circ-arc}.

\begin{corollary}\label{cor:diamondfreeCAhole}
Let $G$ be a diamond\free\ circular-arc graph that contains a hole. If $C = v_1 \ldots v_k$ is a hole of $G$, then the vertices of $G-C$ can be partitioned into $2k$ (possibly empty) pairwise anticomplete sets $U_1,\ldots,U_k,S_1,\ldots,S_k$ such that the following conditions hold.
\begin{itemize}
\item For each $i=1,\ldots,k$, $G[U_i]$ is the disjoint union of cliques and for each $u\in U_i$, $N_C(u)=\{v_i\}$.
\item For each $i=1,\ldots,k$, $G[S_i]$ is a clique and for each $s\in S_i$, $N_C(s)=\{v_i,v_{i+1}\}$.
\end{itemize}
\end{corollary}

\begin{remark}\label{rem:CAdiamond}
In this framework, $G$ is further $\{K_5,H_0\}$\free\ if and only if $|S_i| \leq 2$ for each $i=1,\ldots,k$, the cliques in each $U_i$, $i=1,\ldots,k$, have size at most three, the number of triangles in each $U_i$, $i=1,\ldots,k$, is at most two, and it is at most one if either $S_{i-1}$ or $S_i$ are of size two, and zero if both $S_{i-1}$ and $S_i$ are of size two.
\end{remark}

%--------------------------------------------------------------------------------------------------
We use the following to further simplify the structure of the graphs under consideration.

\begin{lemma}\cite{BMRR-ISCO18}\label{lemma:minimal}
Let $G$ be a $\{K_5,\diamond \}$\free\ graph.\footnote{In \cite{BMRR-ISCO18}, the lemma is stated for chordal graphs but the proof does not use this hypothesis.} If $G$ is a minimal non \BzeroCPG\ graph, then every simplicial vertex of $G$ has degree exactly three.
\end{lemma}

%------------------------------------------------
\begin{figure}[h]
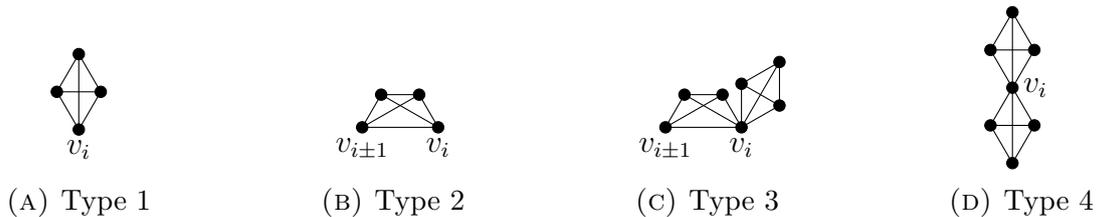

    \centering
    \begin{subfigure}[t]{0.2\textwidth}
        \centering
        \drawingTypeOne
        \caption{Type~1}
    \label{fig:type1}
    \end{subfigure}%
    \hfill
    \begin{subfigure}[t]{0.2\textwidth}
        \centering
        \drawingTypeTwo
        \caption{Type~2}
    \label{fig:type2}
    \end{subfigure}%
    \hfill
    \begin{subfigure}[t]{0.2\textwidth}
        \centering
        \drawingTypeThree
        \caption{Type~3}
    \label{fig:type3}
    \end{subfigure}%
    \hfill
    \begin{subfigure}[t]{0.2\textwidth}
        \centering
        \drawingTypeFour
        \caption{Type~4}
    \label{fig:type4}
    \end{subfigure}
    \caption{Types of a vertex $v_i$ in a hole $v_1, \ldots, v_k$.}
    \label{fig:types}
\end{figure}

In accordance with Corollary~\ref{cor:diamondfreeCAhole}, Remark~\ref{rem:CAdiamond}, and Lemma~\ref{lemma:minimal}, Figure \ref{fig:types} illustrates the different cases that may arise for a vertex $v_i$ in a hole $v_1, \ldots, v_k$ of a $\{K_5,H_0,\diamond\}$\free\ circular-arc graph $G$ which is minimally not \BzeroCPG.
\begin{itemize}
\item Type~0: $U_i=S_i=S_{i-1}=\emptyset$.
\item Type~1: $U_i$ induces a triangle and $S_i=S_{i-1}=\emptyset$.
\item Type~2: $U_i=\emptyset$ and $\max\{|S_i|, |S_{i-1}|\} = 2$.
\item Type~3: $U_i$ induces a triangle and $\max\{|S_i|, |S_{i-1}|\} = 2$.
\item Type~4: $G[U_i]$ is the disjoint union of two triangles.
\end{itemize}

%---------------------------------------------------------------------------------------------------
%---------------------------------------------------------------------------------------------------
% Characterization
\section{Characterisation}
\label{sec:characterization}

%------------------------------------------------
We will call \emph{line} (vertical or horizontal) a 0-bend path on the grid in a \BzeroCPG\ representation of a graph $G$ so as to avoid confusion with paths in $G$. In a \BzeroCPG\ representation of a graph, a \emph{corner} is a point of the grid that belongs to a vertical and a horizontal line.

\begin{lemma}\label{lemma:corners}
The number of corners in a \BzeroCPG\ representation of a hole is even.
\begin{proof}
Let us colour the vertices of the hole according to the representation: a vertex is coloured red (resp. blue) if it is represented by a vertical (resp. horizontal) line. A corner is then determined by two consecutive vertices of the hole that receive different colours. Since the hole starts and ends at the same vertex, and thus, with the same colour, the number of corners is even.
\end{proof}
\end{lemma}

%------------------------------------------------
\begin{lemma}\label{lemma:oddHoleGrid}
Let $C$ be an odd hole. In every \BzeroCPG\ representation of $C$ there are two lines that correspond to consecutive vertices and have the same direction (both vertical or both horizontal).
\begin{proof}
Assume the contrary. Then every pair of consecutive vertices in $C$ determines a corner in its \BzeroCPG\ representation. But the number of pairs of consecutive vertices in an odd hole is odd, which contradicts Lemma~\ref{lemma:corners}.
\end{proof}
\end{lemma}

%------------------------------------------------
\begin{lemma}\label{lemma:contraction}
Let $G$ be a \BzeroCPG\ graph admitting a representation in which the lines $\ell_v$ and $\ell_w$ corresponding to two adjacent vertices $v$ and $w$ have the same direction. Then the graph $G'$ obtained by contracting $v$ and $w$ is also \BzeroCPG.
\begin{proof}
A representation of $G'$ can be obtained by combining $\ell_v$ and $\ell_w$ into a single line.
\end{proof}
\end{lemma}

\begin{corollary}\label{cor:contraction}
Let $C$ be an odd hole of a \BzeroCPG\ graph $G$. Then there are two consecutive vertices of $C$ such that their contraction yields a \BzeroCPG\ graph.
\end{corollary}

%----------------------------------------------------------------
As noticed in previous work \cite{D-G-M-R-cpg,CPGhardness}, any \BzeroCPG\ representation of a $K_4$ necessarily contains a point where coincide one endpoint of each of the lines representing the four vertices. We say that this endpoint of the line is \emph{taken by the} $K_4$, which implies in particular that it cannot be the contact point with a line corresponding to a neighbour outside this $K_4$. It follows that if $\ell$ is a line representing a vertex of Type~4 and $\ell'$ is a line representing one of its neighbour outside the $K_4$s, then the contact point of $\ell$ and $\ell'$ is an interior point of $\ell$ and an endpoint of $\ell'$; in particular, it is a corner.\\

\begin{figure}[h]
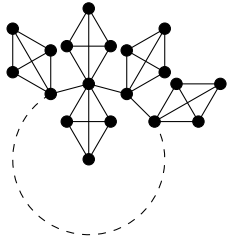

    \centering
    \drawingFtwo
    \caption{$\mathcal{F}_2$: The family of graphs that are an even hole where one of its vertices is of Type~4 and every other vertex of the hole is of Type~1.}
    \label{fig:F2}
\end{figure}

%------------------------------------------------
Let $\mathcal{F}_2$ be the family of graphs that are an even hole where one of its vertices is of Type~4 and every other vertex of the hole is of Type~1 (see Figure~\ref{fig:F2}).

\begin{lemma}\label{lemma:F2}
The graphs in $\mathcal{F}_2$ are not \BzeroCPG.
\begin{proof}
Let $G$ be a graph in $\mathcal{F}_2$. Let $C = v_1 \ldots v_k$ be an even hole of $G$ such that $v_1$ is the vertex of Type~4 and $v_2, \ldots, v_k$ are of Type~1.

Suppose that there is a \BzeroCPG\ representation of $G$ and let $\ell_1, \ldots, \ell_k$ be the lines corresponding to the vertices $v_1, \ldots, v_k$, respectively. Then, every $\ell_i$ with $2 \leq i \leq k$, has one endpoint taken by its corresponding $K_4$, and $\ell_1$ has both endpoints taken. It follows that $\ell_1$ and $\ell_2$ meet at an interior point of $\ell_1$ which is an endpoint of $\ell_2$; and we conclude by induction that for any $i \geq 2$, $\ell_i$ and $\ell_{i+1}$ meet at an interior point of $\ell_i$ which is an endpoint of $\ell_{i+1}$. We then reach a contradiction as $\ell_k$ and $\ell_1$ should meet at an interior point of $\ell_k$ which is an endpoint of $\ell_1$.
\end{proof}
\end{lemma}

\begin{figure}[h]
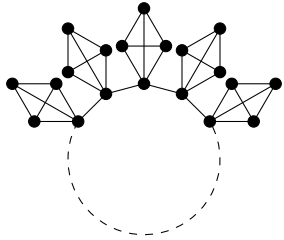

    \centering
    \drawingFthree
    \caption{$\mathcal{F}_3$: The family of graphs that are an odd hole where every vertex of the hole is of Type~1.}
    \label{fig:F3}
\end{figure}

%------------------------------------------------
Let $\mathcal{F}_3$ be the family of graphs that are an odd hole where every vertex of the hole is of Type~1 (see Figure~\ref{fig:F3}).

\begin{lemma}\label{lemma:F3}
The graphs in $\mathcal{F}_3$ are not \BzeroCPG.
\begin{proof}
Let $G$ be a graph in $\mathcal{F}_3$, with odd hole $C$. If $G$ is \BzeroCPG, by Corollary~\ref{cor:contraction}, there are two consecutive vertices of $C$ such that their contraction yields a \BzeroCPG\ graph. But by contracting any two consecutive vertices of $C$ we get a graph in $\mathcal{F}_2$, which is not \BzeroCPG\ by Lemma~\ref{lemma:F2}.
\end{proof}
\end{lemma}

\begin{figure}[h]
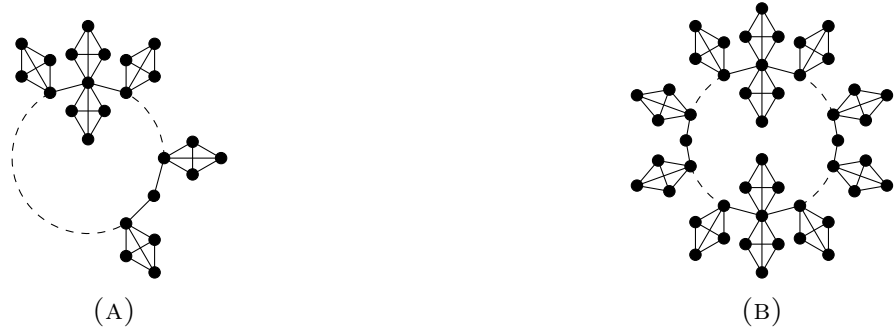

    \centering
    \begin{subfigure}[t]{0.45\textwidth}
        \centering
        \drawingFfourCaseA
        \caption{}
    \label{fig:F4A}
    \end{subfigure}%
    \hfill
    \begin{subfigure}[t]{0.45\textwidth}
        \centering
        \drawingFfourCaseB
        \caption{}
    \label{fig:F4B}
    \end{subfigure}
    \caption{$\mathcal{F}_4$: The family of graphs that are an odd hole containing at least one vertex of Type~4, where  ``between'' every pair of ``consecutive'' vertices of Type~4, there is only one vertex of Type~0 and no vertices of Type~2 nor 3.}
    \label{fig:F4}
\end{figure}

%------------------------------------------------
Let $\mathcal{F}_4$ be the family of graphs that are an odd hole containing at least one vertex of Type~4, where ``between'' every pair of ``consecutive'' vertices of Type~4, there is only one vertex of Type~0 and no vertices of Type~2 nor 3. We say that a pair of vertices $v_i$, $v_j$ (possibly the same) of Type~4 are ``consecutive'' if no vertex in the path $v_{i+1}, \ldots, v_{j-1}$ is of Type~4; and a vertex ``between'' $v_i$ and $v_j$ is any vertex in the path $v_{i+1}, \ldots, v_{j-1}$ (see Figure~\ref{fig:F4}).

\begin{lemma}\label{lemma:F4}
The graphs in $\mathcal{F}_4$ are not \BzeroCPG.
\begin{proof}
It follows from Corollary~\ref{cor:contraction} and the fact that by contracting two consecutive vertices of $C$, we obtain as an induced subgraph either $H_0$, a graph of $\mathcal{F}_1$, or a graph of $\mathcal{F}_2$, which are not \BzeroCPG\ by Lemmas~\ref{lemma:H0K5D}, \ref{lemma:F1}, and~\ref{lemma:F2}.
\end{proof}
\end{lemma}

\begin{figure}[h]
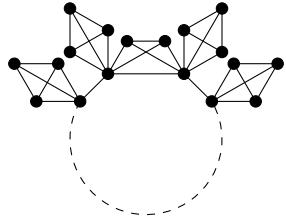

    \centering
    \drawingFfive
    \caption{$\mathcal{F}_5$: The family of graphs that are an even hole where two of its vertices are of Type~3 and all the other vertices of the hole are of Type~1.}
    \label{fig:F5}
\end{figure}

%------------------------------------------------
Let $\mathcal{F}_5$ be the family of graphs that are an even hole where two of its vertices are of Type~3 and all the other vertices of the hole are of Type~1 (see Figure~\ref{fig:F5}).

\begin{lemma}\label{lemma:F5}
The graphs in $\mathcal{F}_5$ are not \BzeroCPG.
\begin{proof}
Let $G$ be a graph in $\mathcal{F}_5$. Let $C = v_1 \ldots v_k$ be an even hole of $G$ such that $v_1$ and $v_2$ are of Type~3, and $v_3, \ldots, v_k$ are of Type~1.

Suppose that there is a \BzeroCPG\ representation of $G$ and let $\ell_1, \ldots, \ell_k$ be the lines corresponding to the vertices $v_1, \ldots, v_k$, respectively. Then, every $\ell_i$ with $3 \leq i \leq k$, has one endpoint taken by its corresponding $K_4$, and $\ell_1$ and $\ell_2$ have a common endpoint  while their other endpoint taken. It follows that $\ell_2$ and $\ell_3$ meet at an interior point of $\ell_2$ which is an endpoint of $\ell_3$; and we conclude by induction that for $i \geq 3$, $\ell_i$ and $\ell_{i+1}$ meet at an interior point of $\ell_i$ which is an endpoint of $\ell_{i+1}$. We then reach a contradiction as $\ell_k$ and $\ell_1$ should meet at an interior point of $\ell_k$ which is an endpoint of $\ell_1$.
\end{proof}
\end{lemma}

%-----------------------------------------------------------------------------------------

Let $G$ be an $H_0$\free\ graph containing a hole $C = v_1 \ldots v_k$, such that the vertices of $G-C$ can be partitioned into $2k$ (possibly empty) pairwise anticomplete sets $U_1,\ldots,U_k,S_1,\ldots,S_k$, where for each $i=1,\ldots,k$ and for each $u\in U_i$, $N_C(u)=\{v_i\}$, and for each $s\in S_i$, $N_C(s)=\{v_i,v_{i+1}\}$; moreover, $G[U_i]$ is either empty, or consists of one or two disjoint triangles; and $G[S_i]$ is either empty or a clique of size two. Notice that the vertices of $C$ can be classified into Type~0, Type~1, Type~2, Type~3, and Type~4. We say that an orientation of some of the edges of $C$ is \emph{feasible} if
\begin{enumerate}
\item\label{it:no-both} no edge is oriented both ways;
\item\label{it:Si} if $S_i \neq \emptyset$ then $v_iv_{i+1}$ is not oriented;
\item\label{it:T4} if $v_i$ is of Type~4, then $v_{i-1}v_i$ and $v_{i+1}v_i$ are oriented this way.
\item\label{it:T3} if $v_i$ is of Type~3 and $S_i \neq \emptyset$ (resp. $S_{i-1} \neq \emptyset$), then $v_{i-1}v_i$ (resp.  $v_{i+1}v_i$) is oriented this way;
\item\label{it:T1} if $v_i$ is of Type~1, then at least one of $v_{i-1}v_i$ and $v_{i+1}v_i$ is oriented this way.
\item\label{it:odd} if $C$ is odd, at least one edge of $C$ is not oriented.
\end{enumerate}

%---------------------------------------------------------------------------------------------------
\begin{lemma}\label{lem:char_order}
Let $G$ and $C$ be defined as above. If $C$ admits a feasible orientation then $G$ is a \BzeroCPG\ graph.
\begin{proof}
The representation of $G$ is based on the ``staircase'' scheme, illustrated in Figure~\ref{fig:stair}, with $\lfloor \frac{k-2}{2} \rfloor$ steps representing vertices $v_1 \ldots v_k$ where the lines in the figure are in clockwise order.

\begin{figure}[h]
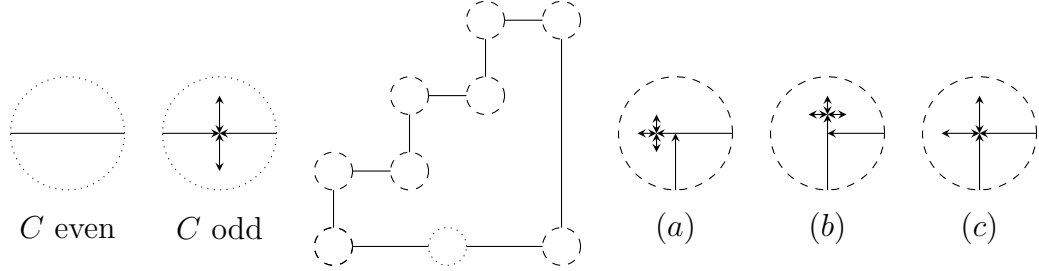

    \centering
    \drawingGridStairCircle
    \caption{Sketch of a staircase \BzeroCPG\ representation of a hole admitting a feasible orientation (the endpoints of a line are marked by an arrow).}
    \label{fig:stair}
\end{figure}

\begin{figure}[h]
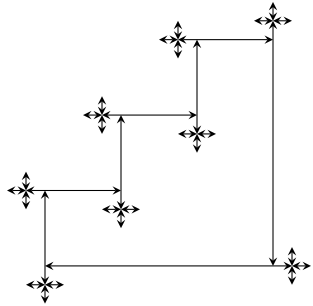

    \centering
    \drawingGridStair
    \caption{Staircase \BzeroCPG\ representation of an even hole where all vertices are of Type~1 and the edges are oriented clockwise.}
    \label{fig:stairT1}
\end{figure}

More specifically, we build a staircase \BzeroCPG\ representation of $G$ given a feasible orientation of $C$, as follows. If $C$ is even, the base of the staircase consists of one line only; and if $C$ is odd, the base of the staircase is formed by two lines corresponding to vertices $v_i$, $v_{i+1}$ such that the edge $v_i v_{i+1}$ is not oriented (in the dotted circle of Figure~\ref{fig:stair} is shown the contact point when $S_i$ is nonempty). For every other $i=1, \dots, k$, the corner formed by the lines corresponding to $v_i$ and $v_{i+1}$ is drawn as shown in the dashed circles of Figure~\ref{fig:stair} (rotated or reflected according to the position of the corner in the staircase), where $(a)$ represents the orientation $v_i v_{i+1}$, $(b)$ represents the orientation $v_{i+1} v_i$, and $(c)$ represents $v_i v_{i+1}$ not oriented. The short lines within the dashed/dotted circles in Figure~\ref{fig:stair} represent the vertices in $U_i$ and $S_i$ that may exist. An example of a staircase \BzeroCPG\ representation is shown in Figure~\ref{fig:stairT1}.
\end{proof}
\end{lemma}

%---------------------------------------------------------------------------------------------------
\begin{theorem}\label{thm:char_hole}
Let $G$ be a circular-arc graph that is not chordal. Let $\mathcal{F} = \mathcal{F}_1 \cup \mathcal{F}_2 \cup \mathcal{F}_3 \cup \mathcal{F}_4 \cup \mathcal{F}_5 \cup \{H_0, \diamond, K_5\}$. Then, $G$ is a \BzeroCPG\ graph if and only if $G$ is $\mathcal{F}$\free.
\begin{proof}
If $G$ is not $\mathcal{F}$\free, then, by Lemmas \ref{lemma:H0K5D}, \ref{lemma:F1}, \ref{lemma:F2}, \ref{lemma:F3}, \ref{lemma:F4} and \ref{lemma:F5}, $G$ is not \BzeroCPG.

Now assume $G$ is $\mathcal{F}$\free\ and let $C = v_1 \ldots v_k$ be a hole of $G$ (which exists as $G$ is not chordal). It then follows from Corollary~\ref{cor:diamondfreeCAhole} that the vertices of $G-C$ can be partitioned into $2k$ (possibly empty) pairwise anticomplete sets $U_1,\ldots,U_k,S_1,\ldots,S_k$ such that for each $i=1,\ldots,k$, $G[U_i]$ is the disjoint union of cliques and for each $u\in U_i$, $N_C(u)=\{v_i\}$; $G[S_i]$ is a clique and for each $s\in S_i$, $N_C(s)=\{v_i,v_{i+1}\}$. Furthermore, by Remark~\ref{rem:CAdiamond}, we have that for each $i=1,\ldots,k$, $|S_i| \leq 2$, and the cliques in $U_i$ have size at most three; moreover, the number of triangles in $U_i$ is at most two, and it is at most one if either $S_{i-1}$ or $S_i$ are of size two, and zero if both $S_{i-1}$ and $S_i$ are of size two. By Lemma~\ref{lemma:minimal}, we may assume henceforth that for each $i=1,\ldots,k$, $|S_i|$ is either zero or two, and that $U_i$ is either empty or the disjoint union of triangles, which allows us to classify the vertices according to their neighbourhood outside $C$ as Type~0, Type~1, Type~2, Type~3, or Type~4.

By Lemma~\ref{lem:char_order}, it suffices to show that $C$ admits a feasible orientation. To this end, consider the connected components of $C$ restricted to the vertices of Type~1.

\textit{Case~1:} Every vertex of $C$ is of Type~1 (the only connected component is a hole).

If $C$ is odd, then $G$ is a graph in $\mathcal{F}_3$, a contradiction. Thus, $C$ is even and orienting every edge as $v_iv_{i+1}$ produces a feasible orientation of the edges of $C$.

\textit{Case~2:} There is only one connected component $P$, which is a path, and only one vertex in $C - P$.

Suppose without loss of generality this vertex is $v_1$. Notice that $v_1$ cannot be of Type~2 or~3 as every vertex of Type~2 or~3 has a neighbour of Type~2 or~3. If $v_1$ is Type~4, then $G$ is either a graph in $\mathcal{F}_2$ or contains a graph in $\mathcal{F}_3$ as induced subgraph, a contradiction. Thus, $v_1$ is of Type~0 and orienting every edge as $v_jv_{j+1}$, for $j=1,\dots,k-1$, while keeping $v_kv_1$ not oriented, produces a feasible orientation of the edges of $C$.

\textit{Case~3:} There is only one connected component $P$, which is a path, and only two vertices in $C - P$.

Notice that these two vertices are necessarily adjacent; thus, we may assume without loss of generality that $v_k$ and $v_1$ are the only two vertices in $C - P$.

If both are of Type~3, then $G$ contains as induced subgraph either a graph in $\mathcal{F}_3$ or a graph in $\mathcal{F}_5$ (according to the parity of $C$), a contradiction. If $v_1$ is Type~3 and $v_k$ is Type~2, a feasible orientation of $C$ is obtained by orienting every edge as $v_{j+1}v_j$, for $j=1,\dots,k-1$, and keeping $v_kv_1$ not oriented (the case where $v_1$ is of Type~2 and $v_k$ is of Type~3 is symmetric). The same orientation remains feasible if $v_1$ and $v_k$ are both of Type~2, although in this case, $v_2v_1$ need not be oriented.

Note that $v_k$ and $v_1$ cannot both be of Type~4 for otherwise they would induce a graph in $\mathcal{F}_1$, a contradiction. Suppose first that one of them is Type~4 and the other Type~0. Then $C$ must be even as $G$ would otherwise be a graph in $\mathcal{F}_4$, a contradiction. Assuming that $v_1$ is of Type~4 and $v_k$ is of Type~0 (the other case is symmetric), a feasible orientation of $C$ is obtained by orienting the edges as $v_{j+1}v_j$, for $j=1,\dots,k-1$, and $v_kv_1$. The same orientation remains feasible if both $v_1$ and $v_k$ are of Type~0, although in this case, edges $v_2v_1$ and $v_kv_1$ need not be oriented (note that at least one of them should not be oriented when $C$ is odd).

\textit{Case~4:} None of the above.

Let $P$ be a (possibly trivial) connected component of $C$ restricted to the vertices of Type~1 (if any). Since we are in neither of the above cases, $P$ is a path and there exist exactly two vertices $u$ and $w$ in $P-C$ having neighbours in $P$. Moreover, $u$ and $w$ are not adjacent. Since $G$ is $\mathcal{F}_1$\free, at least one of them is neither of Type~4 nor of Type~3, say $u$ without loss of generality. Orienting the edge joining $u$ and $P$ towards $P$ and the edges of $P$ in the same direction (clockwise or counter-clockwise), we obtain a partial orientation in which every vertex of $P$ has one incoming edge. By repeating the process for each connected component, we obtain at the end an orientation satisfying the following properties.
\begin{itemize}
\item[--] No edge of $C$ is oriented both ways.
\item[--] No edge of $C$ incident to a vertex of Type~3 or Type~4 is oriented.
\item[--] No edge $v_iv_{i+1}$ of $C$ such that $S_i\neq \emptyset$, is oriented.
\item[--] No edge of $C$ with two endpoints of Type~0 is oriented.
\item[--] Every vertex of Type~1 in $C$ has one incoming edge of $C$.
\end{itemize}

Next, we orient every edge incident to a vertex $v$ of Type~4 towards $v$, and for every vertex $w$ of Type~3 we define the orientation $uw$, where $u$ is the neighbour of $w$ having no common neighbour with $w$. Since $G$ is $\mathcal{F}_1$\free, this orientation is well defined (no edge is incident to two vertices of Types~3 or~4). After this second round of orientation, four of the five properties mentioned above are maintained and the property ``no edge of $C$ incident to a Type~3 or Type~4 is oriented'' is replaced by ``every vertex of Type~3 (resp. Type~4) in $C$ has one (resp. two) incoming edge(s) of $C$.''. A sketch of the orientation process can be found in Figure~\ref{fig:orientation}.

\begin{figure}[h]
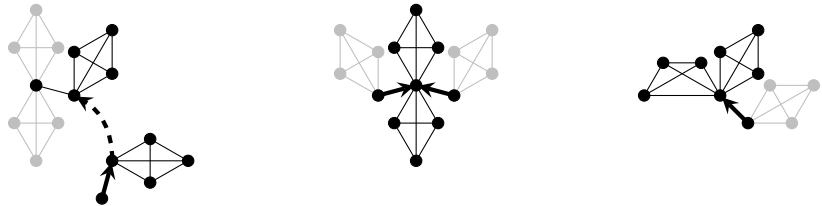

    \centering
    \drawingOrientation
    \caption{Building a feasible orientation of the edges of the hole. Vertices in grey may or may not be present.}
    \label{fig:orientation}
\end{figure}

Thus, in order to ensure that the obtained orientation is a feasible orientation, there remains to show that if $C$ is odd, then there is at least one nonoriented edge. Since this property holds if there are vertices of Type~2 or Type~3, we are left with the case where $C$ odd and only has vertices of Type~4, 1, and~0. Since $G$ is not in $\mathcal{F}_4$, either there exist two adjacent vertices of Type~0 (in which case, the edge joining them is not oriented), or there is a path $P$ of vertices of Type~1 such that the two vertices $u$, $v$ of $C-P$ having neighbours in $P$ are of Type~0. By the rules defined above, none of the edges joining $u$ and $v$ to $P$ was oriented during the second phase, and one of them was left not oriented during the first phase, which concludes the proof.
\end{proof}
\end{theorem}

Combining Corollary~\ref{cor:contact_chordalCA} and Theorem~\ref{thm:char_hole}, we have the following result.

\begin{theorem}\label{thm:charB0CPG_CA}
Let $G$ be a circular-arc graph. Let $\mathcal{F} = \mathcal{F}_1 \cup \mathcal{F}_2 \cup \mathcal{F}_3 \cup \mathcal{F}_4 \cup \mathcal{F}_5 \cup \{H_0, \diamond, K_5\}$. Then, $G$ is a \BzeroCPG\ graph if and only if $G$ is $\mathcal{F}$\free.
\end{theorem}

%---------------------------------------------------------------------------------------------------
%---------------------------------------------------------------------------------------------------
% Algorithm
\section{Algorithm}
\label{sec:algorithm}

In order to recognise the class of \BzeroCPG\ graphs within circular-arc graphs, we first check whether the graph is chordal, which can be done in polynomial time~\cite{R-T-L-chordal}. If it is the case, we can apply the recognition algorithm of~\cite{BMRR-ISCO18}, whose output is either a \BzeroCPG\ representation or a forbidden induced subgraph. Otherwise, we obtain a hole in the graph, and we either find an induced $\diamond$ in the graph or we can compute the structure of the graph with respect to this hole, as described in Corollary~\ref{cor:diamondfreeCAhole}. Each of these steps can be performed in polynomial time.

Once we have computed those different sets, it is easy to check whether it contains either $K_5$ or $H_0$ or none of them. In the latter case, we can use Lemma~\ref{lemma:minimal} to disregard the simplicial vertices of degree one or two, since its proof also suggests how to include them in case we obtain a \BzeroCPG\ representation of the remaining part of the graph.

The remainder of the recognition algorithm is largely based on the proofs of Lemma~\ref{lem:char_order} and Theorem~\ref{thm:char_hole}. We first follow the steps in the proof of Theorem~\ref{thm:char_hole} to either build a feasible orientation of the hole or find a forbidden induced subgraph. In case we obtained a feasible orientation of the hole, we follow the proof of Lemma~\ref{lem:char_order} in order to obtain a \BzeroCPG\ representation of the graph.

%---------------------------------------------------------------------------------------------------
%---------------------------------------------------------------------------------------------------
% Conclusions and further work
%\input{X-Conclusions.tex}

%---------------------------------------------------------------------------------------------------
%---------------------------------------------------------------------------------------------------
% Grants ack
\section*{Acknowledgements}
\label{sec:ack}

This work was done when the second author was visiting the University of Buenos Aires funded by a grant from the Centro Latinoamericano Suizo de la Universidad de San Gallen. It was also partially supported by ANPCyT PICT-2015-2218, and UBACyT Grants 20020170100495BA and 20020160100095BA (Argentina). Carolina L. Gonz\'alez is partially supported by a CONICET doctoral fellowship.

%---------------------------------------------------------------------------------------------------
%---------------------------------------------------------------------------------------------------
\bibliographystyle{plain}
% Styles: plain, acm, abbrv...
% Options: -fl (Name Lastname), -lf (Lastname, Name)

\bibliography{bnm-j,bnm}

%---------------------------------------------------------------------------------------------------
\end{document}